\documentclass{appolb}
\usepackage{graphicx}
\usepackage{authblk}
\usepackage{hyperref}

\begin{document}
\title{ENERGY DEPENDENCE OF FISSION-FRAGMENT NEUTRON MULTIPLICITY IN \textsuperscript{235}U(n,f)
\thanks{Presented at the XXV Nuclear Physics Workshop "Structure and dynamics of atomic nuclei", Kazimierz Dolny, Poland, September 25-30, 2018.}
}
\author{M. ALBERTSSON
\address{Mathematical Physics, Lund University, 221 00 Lund, Sweden}
}

\maketitle

\begin{abstract}
A consistent framework for treating the energy dependence of fission-fragment neutron multiplicities is presented. 
The shape evolution of the compound nucleus towards scission is treated in the strong damping limit using the Metropolis walk method.
The available excitation energy at scission is then divided statistically between the two fragments using microscopic level densities.
Deformation energies, which contribute to the excitation energy when the fragments relax to their ground-state shapes, are also computed. 
From the total fragment excitation energies, the number of emitted neutrons is obtained and illustrated for neutron-induced fission of $^{235}\textrm{U}$.
\end{abstract}
  
\section{Introduction}
Nuclear fission is a complex process and is, despite 80 years after its discovery~\cite{hahn1939,meitner1939}, still not fully understood.
From a theoretical perspective, it is desirable to describe the process within a self-consistent quantum-mechanical microscopic framework.
Although there has recently been a rapid progress in fission models based on density functional theory 
(see \textit{e.g.} Refs.~\cite{regnier2016,bulgac2016,zdeb2017,tao2017}), large-scale calculations with these models require enormous computing time.
To simulate neutron- and $\gamma$-emission one has to employ more phenomenologically-based models. 
Most of the available codes start from mass and total kinetic energy distributions obtained from experiment.
Examples of this are the Los Alamos Model~\cite{madland1982} and its refinements~\cite{madland2017}, 
the PbP model~\cite{tudora2017},
FREYA~\cite{randrup2009}, CGMF~\cite{becker2013} and FIFRELIN~\cite{litaize2010}.
An exception is the semi-empirical GEF model~\cite{schmidt2016} which can calculate quantities for the whole fission process 
but at the expense of introducing several parameters adjusted to experimental data.

Another class of models is based on a classical Langevin description of the fission dynamics 
(see \textit{e.g.} Refs.~\cite{sierk2017,ishizuka2017,sadhukhan2017}), 
which require evaluating a potential energy surface, a collective inertia tensor, and a dissipation tensor.
In the idealized limit of strong dissipation, the shape evolution can be simulated by a Metropolis walk~\cite{randrup2011} on the multi-dimensional 
potential-energy surface.
A consistent framework for treating the energy dependence of the shape evolution was developed~\cite{ward2017} by 
combining the Metropolis walk method with microscopically calculated level densities~\cite{uhrenholt2013}.
The assumption of strongly damped collective motion is also supported by recent calculations based on density functional theory~\cite{stetcu2018}.

We report on further developments~\cite{albertsson2018prl,albertsson2018prc} of the Metropolis walk method 
in which the available excitation energy
at scission is divided between the two nascent fragments based on their microscopically calculated level densities. 
The separated fragments also obtain an additional excitation energy contribution (deformation energy) when they relax to their ground-state shapes.
From the total fragment excitation energies calculated in this way, the number of emitted neutrons from the fragments are obtained and illustrated for 
neutron-induced fission of $^{235}\mathrm{U}$. 

\section{Shape evolution}
The shape evolution is treated by the Metropolis walk method presented in Ref.~\cite{ward2017}.
Formally, this treatment corresponds to the highly dissipative limit of the general Langevin description,
leading to the Smoluchowski equation of motion.
The system is characterized by its shape {\boldmath$\chi$},
described by five parameters: the overall elongation given by the quadrupole moment $Q$, the neck radius $c$,
spheroidal deformations $\varepsilon_{\mathrm{f1}}$ and $\varepsilon_{\mathrm{f2}}$ of the two nascent fragments, and the mass asymmetry $\alpha$. 
The potential energy $U(\mbox{\boldmath$\chi$})$ is calculated within the macroscopic-microscopic model~\cite{moller2009}
in a grid of more than 6 million shapes, where the macroscopic part is calculated within the finite-range liquid-drop model
and the microscopic part is calculated with the folded-Yukawa single-particle model.
The same single-particle model is also employed in the microscopic calculations of level densities~\cite{uhrenholt2013} 
for the 6 million shapes used in the Metropolis walks.
Since most of the structure of the level density appears at low excitation energy,
microscopic level densities are employed up to excitation energy $E^\ast\approx6$ MeV, 
and then smoothly continued upwards by an analytical expression, as described in Ref.~\cite{ward2017}.

Each Metropolis walk is started with a fixed total energy $E_{\mathrm{tot}}$ in the second minimum.
Steps are then taken in the potential-energy landscape based on microscopic level densities
with a local excitation energy given by $E^\ast(\mbox{\boldmath$\chi$})=E_{\mathrm{tot}}-U(\mbox{\boldmath$\chi$})$.
The walks are continued across and beyond the outer barrier until the neck radius $c$ has become smaller than the critical neck-radius $c_0$.
A value of $c_0=1.5$ fm is used based on comparisons with experimental data for the total kinetic energy of the fragments~\cite{albertsson2018prc}.

Once the critical neck-radius has been obtained, the value of the potential energy at this shape $U(\mbox{\boldmath$\chi$})$,
the mass asymmetry coordinate $\alpha$ and the two fragment deformations $\varepsilon_{\mathrm{f1}}$ and $\varepsilon_{\mathrm{f2}}$ are registered.
The number of protons $Z$ and neutrons $N$ in a fragment are determined by requiring the same $Z/N$ ratio as for the fissioning nucleus.
In the present study, only fragments with even mass number and even $Z$ and $N$ are considered.
One million walks are performed to obtain convergence of the results.

\section{Energy partition}
Since the shape evolution is treated in the strong damping limit,
the collective kinetic energy associated with the shape evolution prior to scission is neglected. 
The available excitation energy at scission is then given by the difference between the total energy, $E_{\mathrm{tot}}$, 
and the potential energy of the scission configuration $\mbox{\boldmath$\chi$}_{\mathrm{sc}}$,
\begin{equation}
E^\ast_{\mathrm{sc}}=E_{\mathrm{tot}}-U(\mbox{\boldmath$\chi$}_{\mathrm{sc}}).
\end{equation}
The excitation energy of the compound nucleus at scission is assumed to be divided statistically between the two fragments, \textit{i.e.} 
the heavy-fragment excitation energy, $E^\ast_{\mathrm{H}}$, is given by the following microcanonical distribution:
\begin{equation}
P(E^\ast_{\mathrm{H}},E^\ast_{\mathrm{sc}})\sim\tilde{\rho}(N_{\mathrm{H}},Z_{\mathrm{H}},E^\ast_{\mathrm{H}},\varepsilon_{\mathrm{H}})\cdot
\tilde{\rho}(N_{\mathrm{L}},Z_{\mathrm{L}},E^\ast_{\mathrm{sc}}-E^\ast_{\mathrm{H}},\varepsilon_{\mathrm{L}}),
\end{equation}
where
\begin{equation}
\tilde{\rho}(N_i,Z_i,E^\ast_i,\varepsilon_i)=\sum_{I_i}(2I_i+1)\rho(N_i,Z_i,E^\ast_i,\varepsilon_i,I_i),
\end{equation}
is the effective density of states of a nucleus with neutron and proton numbers $N_i$ and $Z_i$, deformation $\varepsilon_i$, 
and excitation energy $E^\ast_i$, with $i=$ H,L.

In order to take into account the structure effects at low energies,
the same microscopic level density method employed for the compound nucleus is used to calculate level densities 
$\rho(N_i,Z_i,E^\ast_i,\varepsilon_i,I_i)$ of fragments.
All states with different angular momentum $I_i$ are calculated up to excitation energy $E^\ast\approx12$ MeV.
The level densities are then extrapolated to higher excitation energies using Eq.~(8) of Ref.~\cite{ward2017}, 
where the parameter $e_0$ in the formula $a=A/e_0$ is determined by matching with the microscopic level density in the range 9-12 MeV.
Since we are only interested in the energy distribution, we sum over the fragment angular momentum, $I_i$, to obtain
the effective density of states.

\begin{figure}[htb]
\centerline{%
\includegraphics[width=9.0cm]{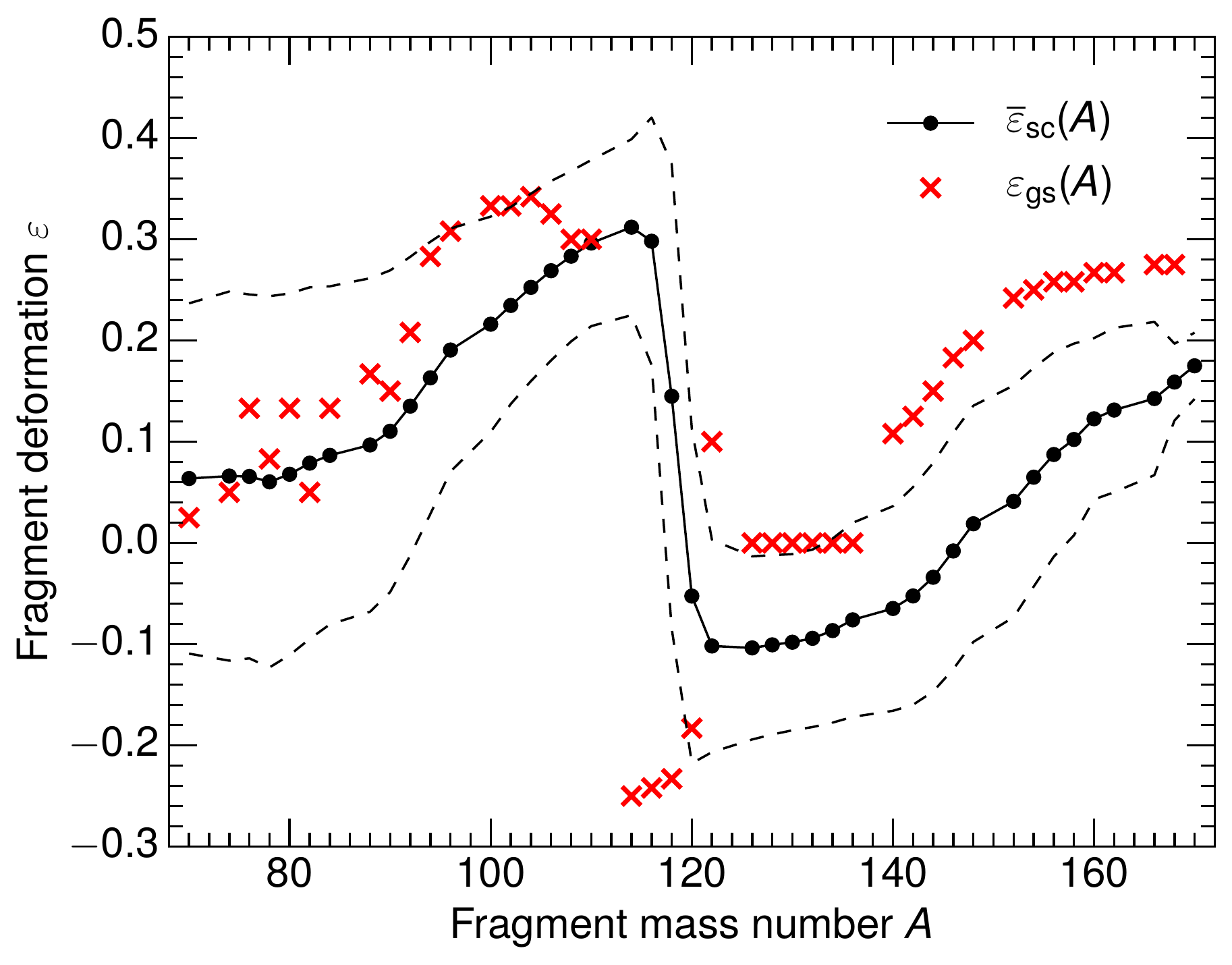}}
\caption{Calculated average fragment deformation at scission (black circles) and standard deviation (black dashed lines) for thermal-neutron induced fission 
of $^{235}\mathrm{U}$ compared to their ground-state deformations (red crosses).
(Fragments $A=114-118$ also have local prolate minima at $\varepsilon\sim0.2$.). Figure from Ref.~\cite{albertsson2018prc}.}
\label{fig:F1H}
\end{figure}

Since the energy is assumed to be partitioned between the fragments at the scission point, 
it is the fragment deformations at scission that need to be used in the calculations of the level densities.
These deformations are usually different compared to the ground-state deformations as seen in Fig.~\ref{fig:F1H}.
Both the calculated average fragment deformations at scission (black circles) and the ground-state deformations (red crosses) 
display a saw-tooth behaviour as a function of fragment mass for thermal-neutron induced fission of $^{235}\mathrm{U}$.
However, the scission deformations tend to be below the values of the ground-states deformations, towards more oblate shapes.
In particular, fragments around $A\sim150$ have deformations at scission furthest away from their ground-state deformations.

\begin{figure}[htb]
\centerline{%
\includegraphics[width=9.0cm]{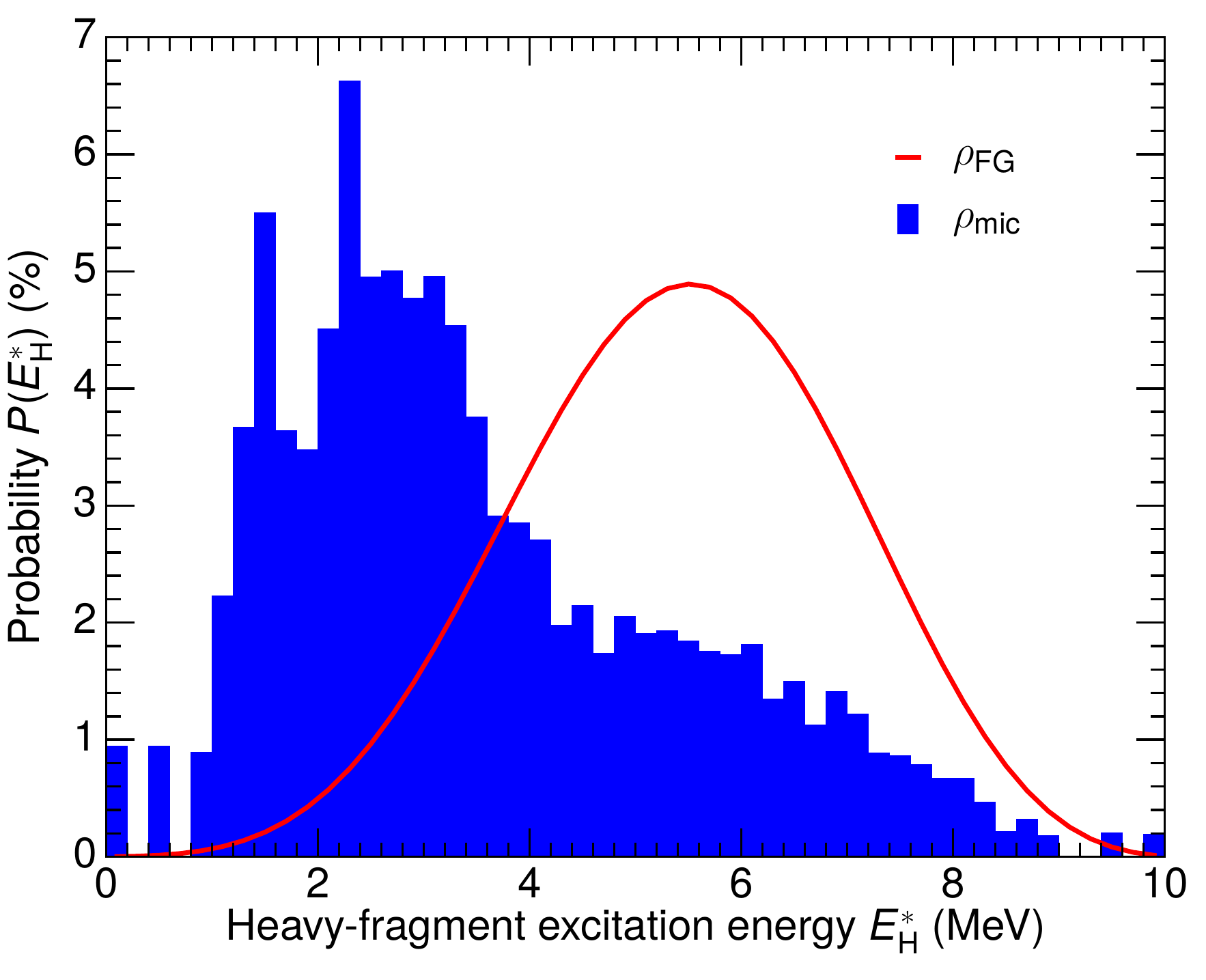}}
\caption{Distribution function $P(E^\ast_{\mathrm{H}})$ for heavy-fragment excitation energy 
for mass-division $(A_{\mathrm{H}}:A_{\mathrm{L}})=(130:106)$ in $^{235}\mathrm{U}(n,f)$ 
for scission energy $E^\ast_{\mathrm{sc}}=10$ MeV,
obtained using microscopic level densities (blue histogram) and Fermi-gas level densities (solid red curve). Figure from Ref.~\cite{albertsson2018prl}.}
\label{fig:F2H}
\end{figure}

It has been seen experimentally that in the split corresponding to $(A_{\mathrm{H}}:A_{\mathrm{L}})=(130:106)$, the heavy fragment emits less neutrons
than the light fragment in thermal-neutron induced fission of $^{235}\mathrm{U}$ (see, \textit{e.g.} Fig.~5 in Ref.~\cite{nishio1998}). 
This is in contrast to expectations if the energy is partitioned based on their heat capacities as in a Fermi-gas model with 
level density $\rho_{\mathrm{FG}}(E^\ast)\sim \mathrm{exp}[2\sqrt{aE^\ast}]$ with $a=A/(8\; \mathrm{MeV})$.
Figure \ref{fig:F2H} shows the energy distribution $P(E^\ast_{\mathrm{H}})$ for the heavy fragment corresponding to this mass-split
for an energy $E^\ast_{\mathrm{sc}}=10$ MeV to be partitioned, and the deformations considered are typical of that division.
Both the energy distribution calculated with the microscopic level density (blue histogram) discussed above and the simplified Fermi-gas level density
(red solid curve) yield rather broad distributions due to the smallness of the nuclear system. 
The macroscopic form yields smooth Gaussian-like distributions peaked at $E^\ast_{\mathrm{H}}/E^\ast_{\mathrm{L}}=A_{\mathrm{H}}/A_{\mathrm{L}}$, 
whereas the microscopic form yields more complicated distributions due to quantal structure effects.

In addition to inherited excitation energy from the compound nucleus, fragments also obtain a deformation energy
$E_{\mathrm{def}}(A)=M(A,\varepsilon_{\mathrm{sc}})-M(A,\varepsilon_{\mathrm{gs}})$,
which is converted into statistical fragment excitations later on as the fragment shapes relax to their ground-state forms.
The shape-dependent fragment masses, $M(A,\varepsilon)$, are calculated in the same macroscopic-microscopic model~\cite{moller2004} 
used in calculations of the potential-energy surfaces.
Fragments $A\sim150$ obtain the largest deformation energy due to having a deformation at scission furthest away from its ground-state deformation
as seen in Fig.~\ref{fig:F1H}.

\section{Neutron multiplicity}
After a fragment has been fully accelerated and its shape has relaxed to its ground-state form, 
it de-excites by emitting neutrons and photons. We only consider neutron emission and assume that no 
photons are emitted until neutron emission are no longer energetically possible.
Since we only consider incident neutrons with low kinetic energy, we also neglect emission of neutrons and photons before scission as well as emission 
from the neck. 

\begin{figure}[htb]
\centerline{%
\includegraphics[width=9.0cm]{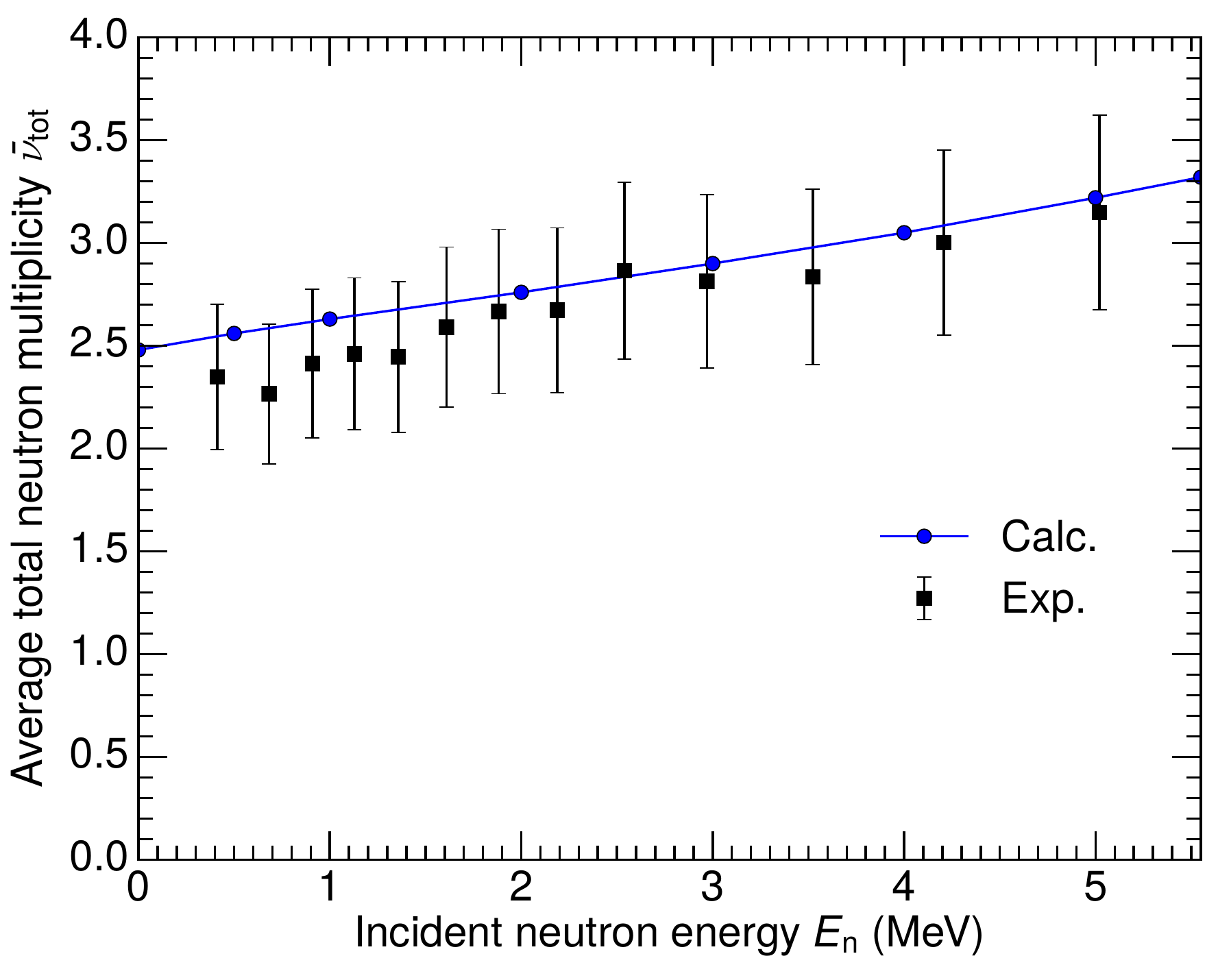}}
\caption{Calculated average total neutron multiplicity $\bar{\nu}_{\mathrm{tot}}$ (blue circles) as a function of the energy of the incident neutron
compared to experimental data~\cite{ethvignot2005} (black squares) for $^{235}\mathrm{U}(n,f)$. Figure from Ref.~\cite{albertsson2018prc}.}
\label{fig:F3H}
\end{figure}

The prompt neutron evaporation is modeled as a single neutron evaporation until no further neutron emission is energetically possible. 
This method was employed in Ref.~\cite{randrup2009} using Fermi-gas level densities. Here, we use microscopically calculated level densities at their
ground-state deformations.
Since the fragment angular momentum $I$ is hardly affected by the evaporation, the energy available for neutron evaporation is taken as
$E=E^\ast-\bar{E}_{\mathrm{rot}}$, where $\bar{E}_{\mathrm{rot}}$ is the average rotational energy (which will later contribute to the photon radiation).
The kinetic energy $\epsilon_{\mathrm{n}}$ of the evaporated neutron from a fragment $(Z,N,E,\varepsilon)$ is sampled from the spectrum
$\sim\tilde{\rho}'(E';\varepsilon')\epsilon_{\mathrm{n}}$, where $\tilde{\rho}'$ denotes the effective level density in the daughter fragment 
$(Z'=Z,N'=N-1,E'=E-\epsilon_{\mathrm{n}}-S_{\mathrm{n}},\varepsilon')$, with $S_{\mathrm{n}}$ being the neutron separation energy in the mother fragment.

\begin{figure}[htb]
\vspace{-0.5mm}
\centerline{%
\includegraphics[width=9.0cm]{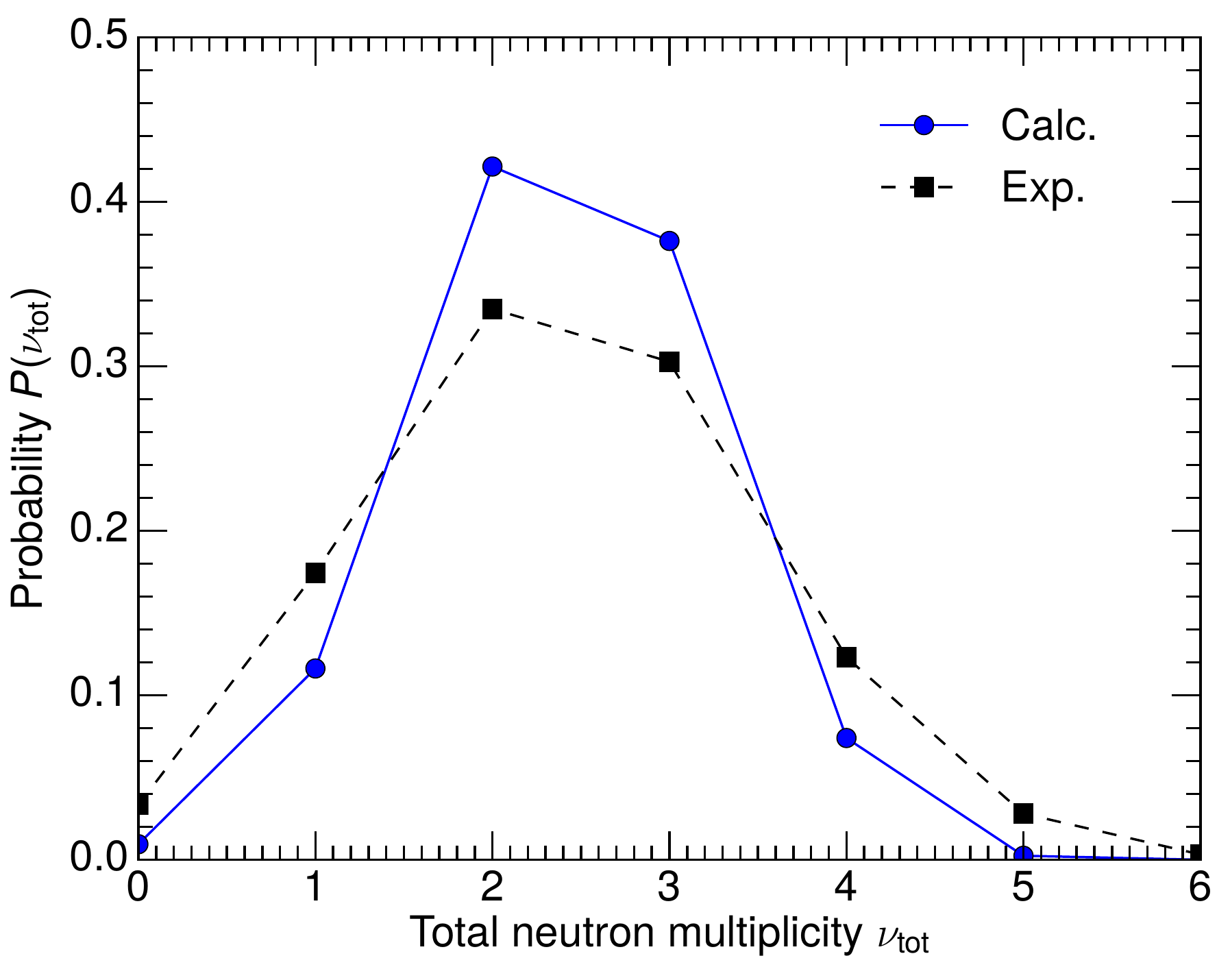}}
\caption{Calculated multiplicity distribution for total emitted neutrons (solid line, blue circles) compared to experimental data~\cite{boldeman1985} 
(dashed line, black squares)
for thermal-neutron-induced fission of $^{235}\mathrm{U}$. Figure from Ref.~\cite{albertsson2018prc}.}
\label{fig:F4H}
\end{figure}

Using the presented model, the neutron multiplicity for each fragment in $^{235}\mathrm{U}(n,f)$ was recently shown~\cite{albertsson2018prl} 
to give good reproduction of experimental data.
The dependence of total neutron multiplicity on incident neutron energy is shown in Fig.~\ref{fig:F3H},
and is also seen to be described well.
In particular, all the calculated values lie inside the experimental error bars.
Figure \ref{fig:F4H} shows the total neutron multiplicity distribution for incident neutron with thermal energy.
The calculated distribution capture the overall behaviour, though the experimental data shows a slightly broader distribution.

\section{Concluding remarks}
We have presented a consistent framework for treating the energy dependence of fission-fragment neutron multiplicities by combining the 
Metropolis walk method with shape-dependent microscopic level densities for the fragments.
Based on Fermi-gas level densities, one would expect the available energy at scission to be divided between the nascent fragments in 
proportion to their heat capacities, while use of microscopic level densities leads to more complicated energy distributions. 
Calculations of the number of emitted neutrons from each fragment with microscopic level densities in 
$^{235}\mathrm{U}(n,f)$ compare well with experimental data~\cite{albertsson2018prl}.

We have shown that the dependence of the average total neutron multiplicity on the energy of the indicent 
neutron also give good reproduction of experimental data.
In addition to average neutron multiplicity, we can describe the width of the distribution reasonably well, though the calculated distribution
is slightly narrower than what is seen experimentally.

The presented model is event-by-event which means that it can describe correlations and fluctuations in quantities, in addition to average values.
It then introduces a considerable predictive power since only potential-energy surfaces and level densities are needed as input.
These are available for all nuclei of interest, and the model can therefore readily be applied to other fission cases as well,
including cases where no experimental data yet exist.

\vspace{5mm}
The presented work was done in collaboration with B.G. Carlsson, T. D{\o}ssing, P. M{\"o}ller, J. Randrup 
and S. {\AA}berg~\cite{albertsson2018prl,albertsson2018prc}.
This work was supported by the Knut and Alice Wallenberg Foundation (Grant No. KAW 2015.0021).


\begin{thebibliography}{9}
\bibitem{hahn1939}
O. Hahn, F. Stra{\ss}mann, \textit{Naturwiss.} \textbf{27}, 11 (1939).

\bibitem{meitner1939}
L. Meitner, O.R. Frisch, \textit{Nature} \textbf{143}, 239 (1939).

\bibitem{regnier2016}
D. Regnier \textit{et al.}, \textit{Phys. Rev. C} \textbf{93}, 054611 (2016).

\bibitem{bulgac2016}
A. Bulgac \textit{et al.}, \textit{Phys. Rev. Lett.} \textbf{116}, 122504 (2016).

\bibitem{zdeb2017}
A. Zdeb, A. Dobrowolski, M. Warda, \textit{Phys. Rev. C} \textbf{95}, 054608 (2017).

\bibitem{tao2017}
H. Tao \textit{et al.}, \textit{Phys. Rev. C} \textbf{96}, 024319 (2017).

\bibitem{madland1982}
D.G. Madland, J.R. Nix, \textit{Nucl. Sci. Eng.} \textbf{81}, 213 (1982).

\bibitem{madland2017}
D.G. Madland, A.C. Kahler, \textit{Nucl. Phys. A} \textbf{957}, 289 (2017).

\bibitem{tudora2017}
A. Tudora, F.-J. Hambsch, \textit{Eur. Phys. J. A} \textbf{53}, 159 (2017).

\bibitem{randrup2009}
J. Randrup, R. Vogt, \textit{Phys. Rev. C} \textbf{80}, 024601 (2009).

\bibitem{becker2013}
B. Becker \textit{et al.}, \textit{Phys. Rev. C} \textbf{87}, 014617 (2013).

\bibitem{litaize2010}
O. Litaize, O. Serot, \textit{Phys. Rev. C} \textbf{82}, 054616 (2010).

\bibitem{schmidt2016}
K.-H. Schmidt \textit{et al.}, \textit{Nucl. Data Sheets} \textbf{131}, 107 (2016).

\bibitem{sierk2017}
A.J. Sierk, \textit{Phys. Rev. C} \textbf{96}, 034603 (2017).

\bibitem{ishizuka2017}
C. Ishizuka \textit{et al.}, \textit{Phys. Rev. C} \textbf{96}, 064616 (2017).

\bibitem{sadhukhan2017}
J. Sadhukhan \textit{et al.}, \textit{Phys. Rev. C} \textbf{96}, 061301 (2017).

\bibitem{randrup2011}
J. Randrup, P. M\"oller, \textit{Phys. Rev. Lett.} \textbf{106}, 132503 (2011).

\bibitem{ward2017}
D.E. Ward \textit{et al.}, \textit{Phys. Rev. C} \textbf{95}, 024618 (2017).

\bibitem{uhrenholt2013}
H. Uhrenholt \textit{et al.}, \textit{Nucl. Phys. A} \textbf{913}, 127 (2013).

\bibitem{stetcu2018}
I. Stetcu \textit{et al.}, arXiv:1810.04024 [nucl-th].

\bibitem{albertsson2018prl}
M. Albertsson \textit{et al.}, submitted to \textit{Phys. Rev. C.}, arXiv:1811.02283 [nucl-th].

\bibitem{albertsson2018prc}
M. Albertsson \textit{et al.}, manuscript in preparation.

\bibitem{moller2009}
P. M\"oller \textit{et al.}, \textit{Phys. Rev. C} \textbf{79}, 064304 (2009).

\bibitem{nishio1998}
K. Nishio \textit{et al.}, \textit{Nucl. Phys. A} \textbf{632}, 540 (1998).

\bibitem{moller2004}
P. M\"oller, A.J. Sierk, A. Iwamoto, \textit{Phys. Rev. Lett.} \textbf{92}, 064304 (2004).

\bibitem{ethvignot2005}
T. Ethvignot \textit{et al.}, \textit{Phys. Rev. Lett.} \textbf{94}, 052701 (2005).

\bibitem{boldeman1985}
J.W. Boldeman, M.G. Hines, \textit{Nucl. Sci. Eng.} \textbf{91}, 114 (1985).

\end{thebibliography}
\end{document}